**Bottom up engineering of near-identical quantum emitters in atomically thin materials**

*Noah Mendelson, Zai-Quan Xu, Toan Trong Tran, Mehran Kianinia, Carlo Bradac, John Scott, Minh Nguyen, James Bishop, Johannes Froch, Blake Regan, Igor Aharonovich\* and Milos Toth\**

*Institute of Biomedical Materials and Devices, University of Technology Sydney, Ultimo, New South Wales, 2007, Australia*
\* igor.aharonovich@uts.edu.au , milos.toth@uts.edu.au

**Quantum technologies require robust and photostable single photon emitters (SPEs) that can be reliably engineered. Hexagonal boron nitride (hBN) has recently emerged as a promising candidate host to bright and optically stable SPEs operating at room temperature. However, the emission wavelength of the fluorescent defects in hBN has, to date, been shown to be uncontrolled. The emitters usually display a large spread of zero phonon line (ZPL) energies spanning over a broad spectral range (hundreds of nanometers), which hinders the potential development of hBN-based devices and applications. We demonstrate bottom-up, chemical vapor deposition growth of large-area, few-layer hBN that hosts large quantities of SPEs: ~100 per 10 × 10 μm$^2$. Remarkably, more than 85% of the emitters have a ZPL at (580 ± 10) nm—a distribution which is over an order of magnitude narrower than previously reported. Exploiting the high density and uniformity of the emitters, we demonstrate electrical modulation and tuning of the ZPL emission wavelength by up to 15 nm. Our results constitute a definite advancement towards the practical deployment of hBN single photon emitters in scalable quantum photonic and hybrid optoelectronic devices based on 2D materials.**

Solid-state, single-photon emitters (SPEs) are highly attractive for scalable quantum photonic applications, in particular for quantum cryptography, quantum repeaters and entanglement distribution.[1-6] In the past few years, major advances have been achieved in the characterization of single defects in wide-bandgap materials,[7-9] as well as single molecules,[10] quantum dots[1] and, more recently, layered van der Waals materials[11-15] —which resulted in a vast library of available SPEs. However, one of the overarching challenges across all platforms is the inability to readily engineer large area (centimeter-scale) materials with uniform spatial and spectral distributions of SPEs. Such a platform is highly desirable in the context of developing future scalable quantum photonic architectures.

To tackle this challenge, two-dimensional (2D) materials which can be fabricated by chemical vapour deposition (CVD) are highly advantageous.[16] They can be grown at low cost, using conventional furnaces and precursors, over large areas. However, to date, materials synthesized via this method have not been shown to possess densities of SPE high enough to be desirable for scalable, integrated technologies. The fabrication of quantum emitter arrays has been pursued by positioning exfoliated flakes of transition metal dichalcogenides (TMDCs) on top of lithographically-predefined pillars.[17, 18] This process is expensive as it requires e-beam lithography and a subsequent reactive ion-etching step, and the size of the 2D material is limited by the exfoliation process. Moreover, this method has only worked reliably with TMDCs, for which the fabricated SPEs operate at cryogenic temperatures and cannot be transferred to different substrates for device fabrication, since the SPEs require activation via the strain field induced by the pillars.

Similar attempts with other 2D materials (e.g. hBN) have so far been shown to be inconclusive.[19, 20]

Here we demonstrate that all these limitations can be overcome by bottom-up fabrication of hBN using low-pressure chemical vapor deposition (LPCVD). The technique yields large, centimeter-scale, high-quality hBN that hosts a high density of bright SPEs. The method stands out in terms of scalability and low cost—which are highly desirable for the engineering of integrated devices based on SPEs in a 2D material.[21-23] More importantly, for the first time, the as-grown hBN SPEs exhibit a narrow, uniform distribution of ZPLs that can be tuned electrically, solving a critical bottleneck in the practical implementation of quantum technology.

Figure 1a shows a schematic illustration of our growth setup (cf. Methods for details). Briefly, the growth was performed in a quartz tube furnace using ammonia borane ($BH_3NH_3$) as a precursor, at a temperature of 1030 C, with $Ar/H_2$ as a carrier gas. These conditions yield high quality, uniform hBN films (typically less than 10 monolayers) over length scales greater than a $cm^2$. An optical image of a hBN film on a copper substrate is shown in Figure 1b, and a scanning electron microscope (SEM) image of the hBN film is shown in Figure 1c. Periodic wrinkles and grain boundaries are visible—which is typical of hBN grown by CVD. The thickness of the films was determined by AFM (Fig. 1e) and shown to be 1.4 nm for this particular growth, corresponding to a stack of three-to-four monolayers of hBN. The high quality of the as-grown hBN was confirmed by Raman (Fig. 1d) and Fourier-transform infrared (FTIR) spectroscopy (cf. Supporting Information). A Raman peak centered at 1369 $cm^{-1}$ is clearly visible, which is characteristic of the $E_{2g}$ mode of few-layer hBN.

To perform further optical characterization, the as-grown hBN was transferred to a $Si/SiO_2$ substrate using a polymer-assisted wet transfer technique (cf. Methods). When present, the hydrocarbon residues remaining after transfer were removed via additional annealing in air at 550 C for 120 minutes. In contrast to prior reports,[13-15, 24, 25] post-growth annealing was not necessary to activate or stabilize the quantum emitters. The optical properties of the emitters were studied at room temperature using a lab-built scanning confocal microscope with a 532-nm excitation laser (cf. Methods). The collected emission was filtered using a 568-nm long pass filter, allowing for the observation of hBN emitters at longer wavelengths. Figure 2a displays an histogram of the emission wavelength values measured for 248 individual narrow-band emitters. The inset shows PL spectra from a subset of seven single emitters with ZPLs in the range ~575–590 nm. All emitters display a phonon sideband approximately 165 meV below the ZPL energy. More than 85% of the emitters were measured to have a ZPL wavelength localized in a very narrow spectral window, (580 ± 10) nm, and up to 56% in the narrower spectral window (580 ± 5) nm. Such a narrow wavelength distribution constitutes a significant advance in the spectral localization of hBN defect emission, which across the literature has been usually reported to span over two hundred nanometers.[14, 15, 26] Importantly, the obtained narrow distribution is consistently reproducible over multiple growths. Figure 2b shows similar ZPL wavelength distributions—centered at 580 nm—of single emitters measured in hBN samples fabricated in five separate growths, using identical conditions. In general, we employ up to five copper substrates simultaneously in each growth, each substrate is ~1 $cm^2$, and we typically achieve full surface coverage and uniformity in all resulting hBN films. The relatively high fabrication yield, coupled with the reliability of transfer techniques for CVD-grown 2D materials gives hBN a distinct advantage over other material platforms identified as potential hardware for scalable quantum photonic devices.

In addition to the narrow ZPL energy distribution, the density of emitters in the as-grown films is high and uniform. This is critical for device scalability and integration. For instance, the high density of QDs (>10 per µm$^2$) in GaAs/InGaAs wafers has enabled coupling of single dots to photonic crystal cavities owing to the high probability of spatial overlap between an emitter and a fabricated cavity.[27] In our experiments, the emitter density was probed by confocal photoluminescence (PL) imaging. Figure 3a is a 5×5 µm$^2$ confocal map of the sample. Red and green circles indicate the locations of ensembles and single emitters, respectively (classified using autocorrelation measurements). Our results show, across all characterized samples, an overall density of 2.2 emitters/µm$^2$ and a single emitter density of 0.84 emitters/µm$^2$ (Fig. 3b). Both values constitute significant improvements over previously published results on emitter densities in hBN,[24, 25] including in commercially-available sources, as evidenced by wide-field EMCCD microscopy measurements shown in Figure S5. The quantum nature of SPEs was confirmed by recording the second-order autocorrelation function, $g^2(\tau)$. Figure 3c shows representative $g^2(\tau)$ measurements for twelve emitters in the area mapped in Figure 3b. In all cases, the intensity at zero-delay time, $g^2(\tau = 0)$, is below 0.5, confirming the presence of single quantum emitters. Note that the $g^2(\tau)$ curves in Figure 3c are not background-corrected to showcase the high purity of the SPEs and their suitability for practical devices. Overall, the emitters are bright with saturation count rates exceeding $1 \times 10^6$ counts/s. The majority of the SPEs are optically stable and do not undergo blinking (approximately 15% exhibit blinking and less than ~5% photobleach entirely during PL excitation). More examples and a detailed analysis of the photophysical properties of the emitters are shown in the Supporting Information (Fig. S4).

Remarkably, the emitters in the LPCVD-grown hBN films are not concentrated preferentially or exclusively at the grain boundaries or other large-scale defects—a characteristic which is instead typical of exfoliated hBN flakes.[24, 28, 29] This is a substantial advantage for incorporation of the emitters in devices. Furthermore, the narrow ZPL energy distribution suggests that the growth conditions favour the incorporation of a particular type of defects—in contrast to the majority of other hBN samples studied to date, which exhibit broad ZPL energy distributions and are likely the result of multiple defect species. The chemical/atomic structure of the emitters is, in fact, still unknown and a matter of much debate in the field.[30-32] Samples grown using LPCVD therefore offer a promising avenue to shine new light on the problem, both because the SPE energy distribution in our nominally-undoped samples is narrow, and because the dopants and growth conditions that affect defect formation energies are easy to control during LPCVD growth.

Regardless, our LPCVD material already surpasses many of the practical limitations of other available hBN sources which are currently hindering the realization of integrated photonic devices. For instance, commercially-available hBN material in the form of small flakes ~200 nm in size—currently the most studied hBN samples[13, 14, 26]—are usually obtained by liquid exfoliation with metal ions and drop-casted onto substrates, which results in random dispersion. Similarly, tape-exfoliated hBN from crystal sources results in unreliable thicknesses and highly nonuniform emitter distributions that are concentrated at flake edges, wrinkles and grain boundaries—neither is suitable for device integration. In contrast, the LPCVD process yields hBN with controllable thickness and a uniform distribution of emitters over large length scales.

To further advance the functionality of our synthesized hBN, we combine it with a polymer electrolyte (PEO:LiClO$_4$, cf. Methods) to electrically modulate the emission of the hosted SPEs using the device configuration shown in Figure 4a. This polymer electrolyte is widely used for on-chip nanophotonic and opto-electronic devices with exceptional performance, owing to a strong electric field deriving from a double-layer of charges on the surface.[33] Applying a negative voltage to the hBN electrode draws Li$^+$ to the electrode/hBN interface to form a double layer of opposing charges between the polymer and the electrode. Since the distance between the charges is ~1 nm, the resulting capacitance is extremely high, ~2200 nF/cm$^2$, (a 300-nm-thick SiO$_2$ based back gate gives a capacitance of ~10 nF/cm$^2$), which enables efficient control of the electric field intensity using a bias of only a few volts.

The as-grown hBN film was transferred onto a lithographically-defined gold electrode separated (by ~15 μm) from a second reference electrode (see Fig. 4a). We identified SPEs within the hBN and then applied a voltage to tune their wavelengths. Figure 4b shows a typical SPE with a ZPL at ~580 nm undergoing a voltage cycle from 0 to ± 6 V. A shift of ~2.7 nm in the ZPL wavelength is observed for voltage values of ± 6 V; the return of the ZPL to the initial spectral position after the voltage is turned off shows that the process is reversible. Notably, the emission undergoes a red shift under both positive and negative bias. Figure 4c shows a PL spectrum of an SPE with a ZPL at ~622 nm acquired under biases spanning ±6 V. The PL was collected continuously every 10 s for 200 s for each value of the voltage, and no spontaneous spectral jumps were observed. When the voltage was gradually changed to –6V the ZPL blue-shifted to ~615 nm, while when the voltage was switched to +6V the ZPL red-shifted to ~630 nm. Overall, this SPE exhibited a continuous ZPL shift of over 15 nm (~50 meV) when the voltage was changed from –6 V to +6 V. Such a modulation of a SPE wavelength is the largest obtained to date from any solid state SPE, or from the excitonic transition of a 2D material.[34-36] We note that the sensitivity of the emission wavelength to the applied bias varied for different SPEs, likely due to variations in dipole orientation with respect to the electrodes, and variations in surrounding charge traps, and local dielectric environment. Nonetheless, our results provide a promising way to tune the emission wavelength.

In summary, we have demonstrated a robust method for the production of centimeter-scale, few-layer hBN, with stable, bright and high-quality SPEs. The emitters are uniformly distributed in an almost atomically-thin layer of hBN, and exhibit emission lines predominantly at (580 ± 10) nm, thus solving one of the most critical bottlenecks in the field of solid-state quantum photonics. The presence of uniform emitters within the flakes will enable faster and easier integration with photonic devices. Furthermore, having a high density of hBN emitters with uniform properties will expedite critical experiments such as two-photon interference and indistinguishability. One particularly appealing aspect of our work is the implementation of LPCVD hBN emitters as a platform for multiplexing technologies requiring multiple SPEs operating within a narrow but not identical spectral range, to yield multichannel narrowband quantum device integration.[37] Finally, we also show that by employing a polymer electrolyte, tuning of the emission wavelength of hBN is possible, achieving a record shift for an SPE of up to 15 nm (~ 50 meV). Such tuning can be further explored to understand the crystallographic nature of SPEs in hBN, and pave the way towards the deterministic engineering of these emitters in bulk hBN crystals. Overall, our work provides important insights into designing scalable nanophotonic devices employing 2D materials, in particular hBN.

**Acknowledgements**
Financial support from the Australian Research council (via DP180100077), the Asian Office of Aerospace Research and Development grant FA2386-17-1-4064, the Office of Naval Research Global under grant number N62909-18-1-2025 are gratefully acknowledged. This research is supported by an Australian Government Research Training Program Scholarship. The authors thank Sejeong Kim and Alex Solntsev for useful discussions.

**Figures**

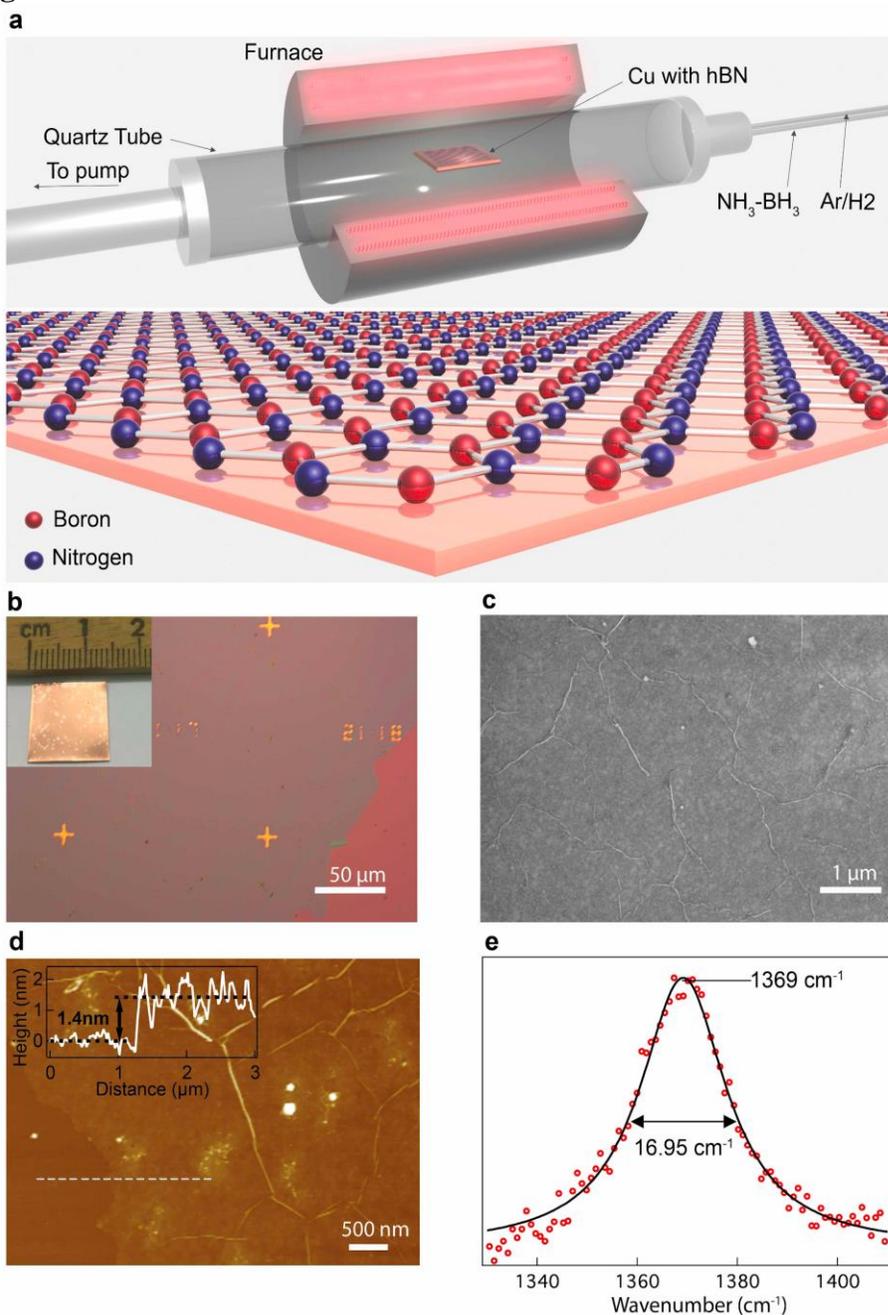

*Fig. 1 Growth and properties of CVD hBN films.* ***a.*** *Schematic of LPCVD growth of hBN film on copper. Ammonia borane is used as a CVD precursor.* ***b.*** *Optical image of the as-grown hBN film transferred onto a Si/SiO$_2$ substrate marked with gold numbers, showing uniformity of the film. Inset: photograph of the copper foil substrate prior to hBN transfer, showing the cm$^2$-scale growth area.* ***c.*** *SEM image showing the morphology of the hBN domains transferred onto the Si/SiO$_2$ substrate.* ***d.*** *AFM image of the hBN film on the Si/SiO$_2$ substrate. Inset: height profile of hBN film edge, showing a 1.4-nm step, corresponding to ~3 atomic layers of hBN.* ***e.*** *Raman spectra of the hBN film displaying the characteristic E$_{2g}$ mode measured using an excitation wavelength of 633 nm.*

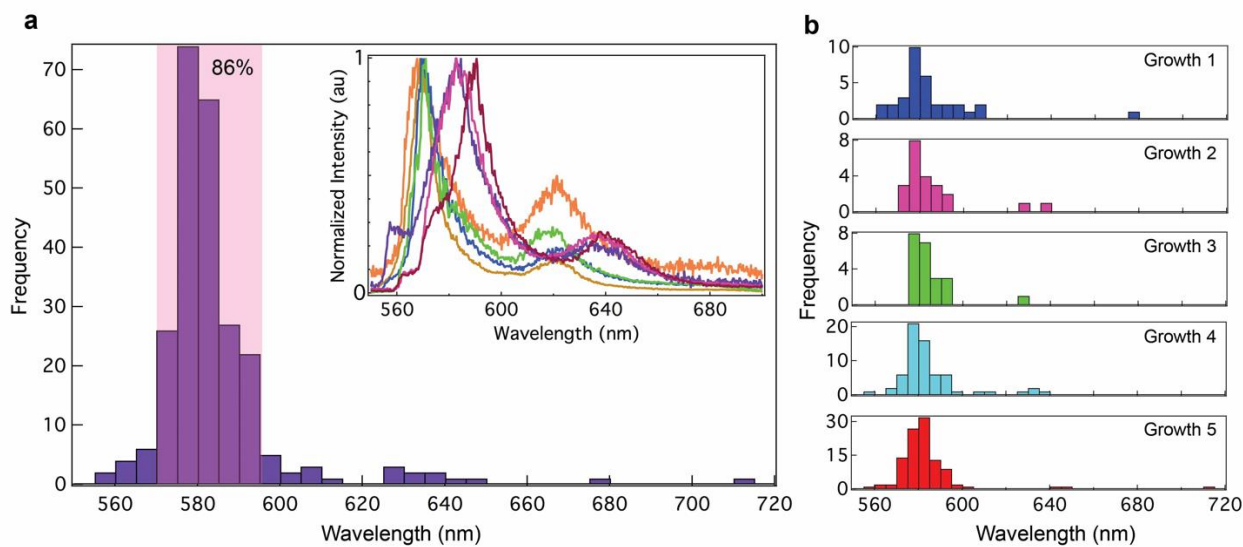

*Fig. 2 Spectral distribution of emitters in CVD-grown hBN. **a.*** *Histogram of hBN ZPL wavelengths produced using a bin size of 5 nm. A total of 248 emitters from five different hBN films (i.e., growths) were used to generate the histogram. The shaded area highlights the range $\lambda_{emisson}= (580 \pm 10)$ nm, which contains 86% of the emitters. Inset: characteristic room-temperature emission profile of 7 emitters from the same spectral region, all showing a typical phonon sideband of ~165 meV. **b**. Individual ZPL distribution histograms from five different growths with nominally identical experimental parameters, showing the robust and reproducible ZPL localization achieved by the LPCVD growth technique (bin size = 5 nm).*

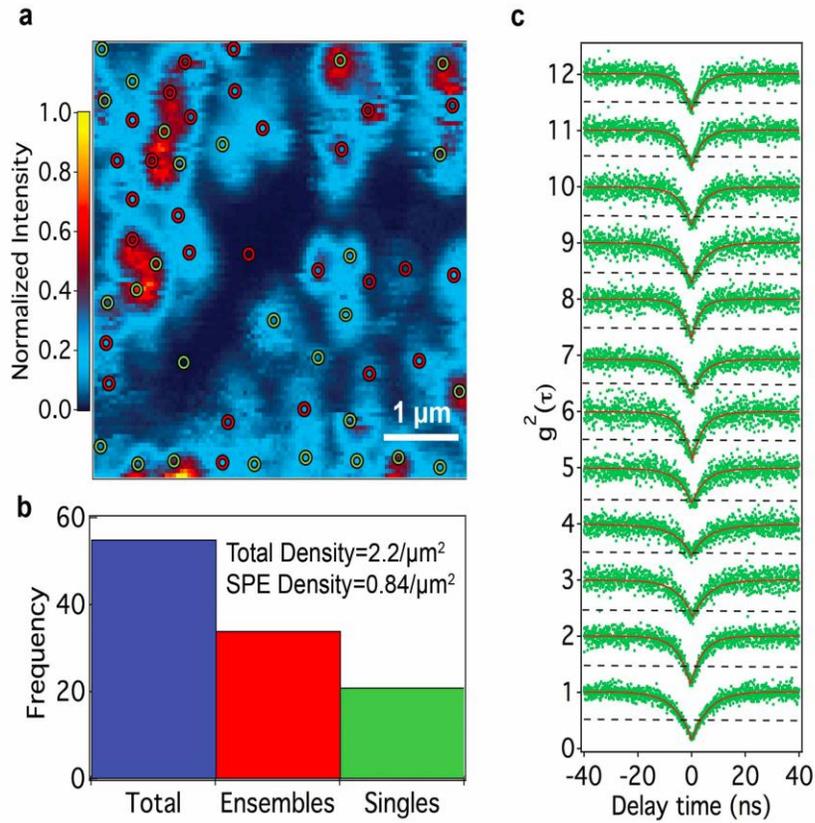

*Fig. 3 Single photon emitter density. a.* Confocal photoluminescence map of a 5×5 μm$^2$ area where 55 emitters were characterized. *b.* Histogram of the characterized emitters displaying the total (blue bar), ensemble (red bar), and single (green bar) photon emitter occurrence frequencies found in the confocal scan. The overall emitter densities are ~2 emitters/μm$^2$ and ~1 single emitters/μm$^2$. *c.* Characteristic autocorrelation measurements of twelve SPEs within the scanned region: the dashed lines indicate the 0.5 threshold ($g^2(0) \leq 0.5$) for single photon emitters. The curves are not corrected for background luminescence. The plots are offset vertically for display purposes.

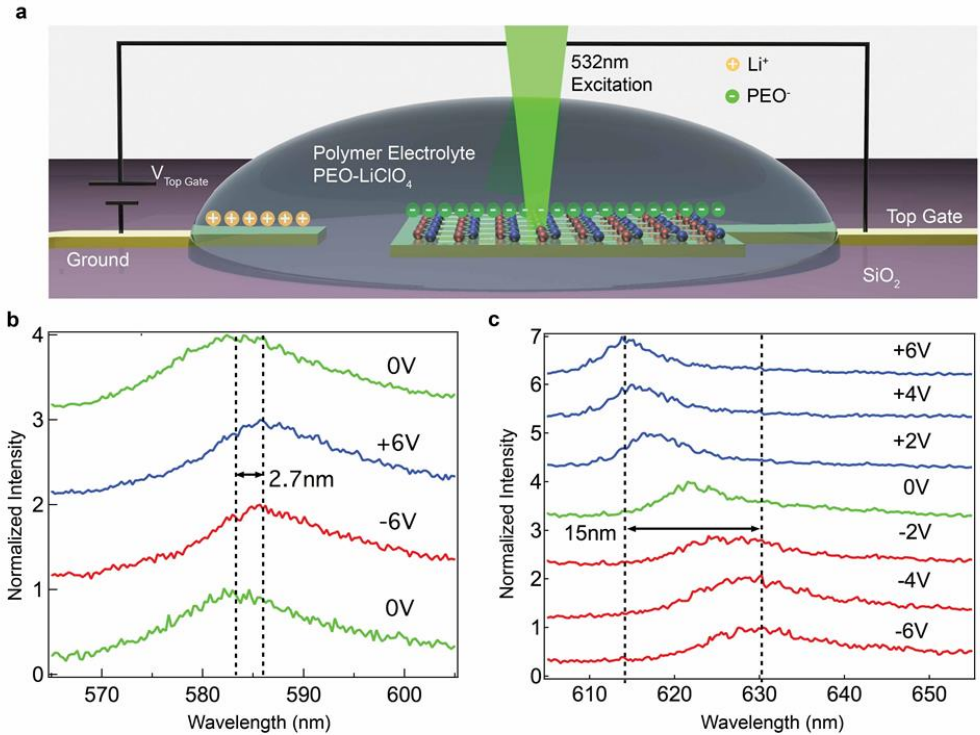

*Fig. 4 Spectral modulation of hBN SPEs. **a.** Schematic of a polymer electrolyte device used to modulate the hBN emission. CVD-grown hBN was transferred onto one of the gold electrodes, and a polymer electrolyte was spin-coated onto the device. **b.** Modulation of the SPE with a ZPL at 585 nm, displaying a 2.7 nm shift upon the application of a ± 6V bias. **c.** Modulation of a SPE which can be tuned by up to 15 nm using a bias voltage in the range of ±6 V.*